\documentclass[useAMS,usenatbib,usegraphicx]{mn2e}

\def\kms{km s$^{-1}$} \def\msun{M$_{\sun}$} \def\rsun{R$_{\sun}$} \def\aap{A\&A}
\def\apjl{ApJ} \def\apj{ApJ}  \def\aj{AJ} \def\mnras{MNRAS}
 \def\pasp{PASP} 
\newcommand\ion[2]{#1$\;${\scshape{#2}}}

\title[AM CVn Progenitors]{Found: The Progenitors of AM CVn and Supernovae .Ia}

\author[M. Kilic et al.]
{
Mukremin Kilic$^1$\thanks{kilic@ou.edu},
J. J. Hermes$^{2,3}$,
A. Gianninas$^1$,
Warren R. Brown$^4$,
Craig O. Heinke$^5$,
\newauthor
M. A. Ag\"{u}eros$^6$,
Paul Chote$^{7,8}$,
Denis J. Sullivan$^{7,8}$,
Keaton J. Bell$^2$,
Samuel T. Harrold$^2$\\
$^1$Department of Physics and Astronomy, University of Oklahoma, 440 W. Brooks St., Norman, OK, 73019, USA\\
$^2$Department of Astronomy, University of Texas at Austin, RLM 16.236, Austin, TX 78712, USA\\
$^3$Department of Physics, University of Warwick, Coventry CV4 7AL, United Kingdom\\
$^4$Smithsonian Astrophysical Observatory, 60 Garden St, Cambridge, MA 02138, USA\\
$^5$Department of Physics, CCIS 4-183, University of Alberta, Edmonton, AB, T6G 2E1, Canada \\
$^6$Department of Astronomy, Columbia University, 550 West 120th Street, New York, NY 10027, USA \\
$^7$School of Chemical \& Physical Sciences, Victoria University of Wellington, New Zealand \\
$^8$Visiting Astronomer, Mt. John University Observatory, operated by the University of Canterbury, New Zealand
}

\begin{document}

\maketitle

\begin{abstract}
We present optical and X-ray observations of two tidally distorted, extremely low-mass white dwarfs (WDs)
with massive companions. There is no evidence of neutron stars in our {\em Chandra} and {\em XMM}
observations of these objects. SDSS J075141.18$-$014120.9 (J0751) is an
eclipsing double WD binary containing a 0.19\msun\ WD with a 0.97\msun\
companion in a 1.9 h orbit. J0751 becomes the fifth eclipsing double WD system currently known. 
SDSS J174140.49+652638.7 (J1741) is another binary containing a 0.17\msun\ WD
with an unseen $M\geq1.11$\msun\ WD companion in a 1.5 h orbit. With a mass ratio of $\approx0.1$, J1741 will have
stable mass transfer through an accretion disk and turn into an interacting AM Canum Venaticorum
(AM CVn) system in the next $\approx$160 Myr. With a mass ratio of 0.2, J0751 is likely
to follow a similar evolutionary path. These are the first known AM CVn progenitor binary
systems and they provide important constraints on the initial conditions for AM CVn.
Theoretical studies suggest that both J0751 and J1741 may create thermonuclear supernovae in $\sim10^8$ yr,
either .Ia or Ia. Such explosions can account for $\sim$1\% of the Type Ia supernova rate.
\end{abstract}


\section{INTRODUCTION}

AM CVn are interacting binary systems involving an accreting white dwarf (WD)
and a helium-rich donor star, which may be either low-mass WDs, helium stars, or
evolved main-sequence stars. There are currently more than 30
AM CVn stars known with orbital periods ranging from 5.4 min \citep{roelofs10} to 65 min
\citep{levitan13,carter13}. 
The prototype of the class, AM CVn, was discovered by \citet{smak67}, but
the relative importance of the proposed birth channels is still not well understood.
The problem is that all three channels lead to the same donors: very low-mass ($<0.1$\msun)
degenerate dwarfs \citep{nelemans10}. Studying the chemical abundances of 11 AM CVn systems,
\citet{nelemans10} find evidence for WD donors in three systems, but the donor type is not clear
for the remainder of their sample.

The orbital evolution of AM CVn is initially dominated by gravitational wave radiation.
However, after the mass transfer starts to dominate the evolution, the orbital period goes through
a minimum and then increases \citep{solheim10}.
For the shortest-period AM CVn, the mass transfer rate is high enough that the accreted helium onto the
CO WD undergoes unstable, nova-like outbursts. As the orbital period increases the mass transfer rate decreases, and the required mass for unstable burning goes up, leading to a final
flash with a helium layer mass of $\sim$$0.02-0.1$ \msun. This occurs within $<10^8$ yr of
reaching contact. \citet{bildsten07} find that the final
flash for a $>$ $0.7-0.9$ \msun\ CO WD is likely to be dynamical, leading to the ejection of radioactive
$^{48}$Cr, $^{52}$Fe, and $^{56}$Ni. This might lead to a faint and rapid thermonuclear supernova (SN) .Ia.
Such a transient event might have already been observed \citep{kasliwal10}.

Given the difficulties in disentangling the component masses in AM CVn,
the identification of their progenitor systems would be extremely useful for constraining the
initial conditions for AM CVn and possible SNe .Ia explosions.
Here we present the identification of two such progenitor systems, J0751 and J1741.
Both were discovered in the ELM (Extremely Low Mass) Survey \citep{brown13,kilic12} as
short period binaries with massive companions. We obtained follow-up optical
and X-ray observations to constrain the nature of the ELM WDs and their unseen companions.
Our observations are discussed in \S 2,
and the binary parameters are discussed in \S 3. Evolutionary scenarios for
the future of these systems are presented in \S 4. 

\vspace{-0.2in}
\section{Observations}

\subsection{Optical Spectroscopy}

\begin{figure}\vspace{-1.5in}\hspace{-0.35in}
\includegraphics[width=4.0in,angle=0]{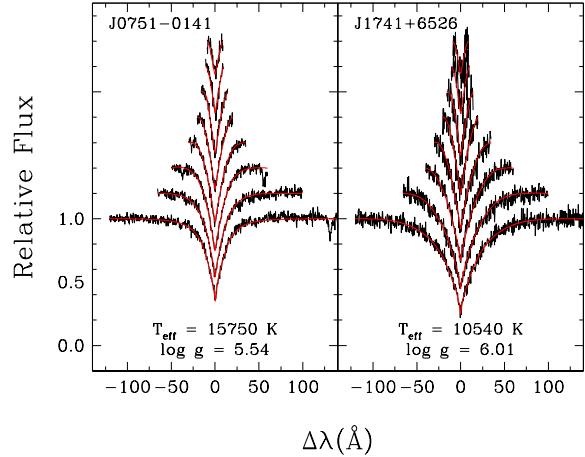}
\vspace{-1.6in}
\caption{Model fits (red lines) to the Balmer line profiles of J0751 and J1741.
H$\gamma$ (bottom) through H12 (top) are shown.
\label{fig:spec}}
\vspace{-0.1in}
\end{figure}

Optical spectroscopy of the ELM WDs in J0751 and J1741 were previously presented by
\citet{brown12,brown13}. J1741 was originally found to be one of the lowest surface
gravity WDs known, with a spectroscopically determined $\log{g}<5.2$. However, this
low surface gravity is incompatible with the relatively small inferred radius of the ELM 
WD determined from the ellipsoidal variations observed in its optical light curve
\citep{hermes12}. We used an extended model atmosphere grid based on the \citet{tremblay09}
models to re-evaluate the spectroscopic temperature and surface gravity for both J0751 and
J1741. Figure \ref{fig:spec} shows our model fits to the Balmer line profiles of both stars.

The best-fit model for J0751 has $T_{\rm eff} = 15750 \pm 240$ K and $\log{g} = 5.54 \pm 0.05$.
The recent evolutionary calculations by \citet{althaus13} demonstrate that
J0751 is a $280 \pm 140$ Myr old, 0.194(6)\msun, and $M_g=6.1$ mag WD, at a distance of $1.7 \pm 0.1$ kpc.
The best-fit model 
for J1741 has $T_{\rm eff} = 10540 \pm 170$ K and $\log{g} = 6.01 \pm 0.06$.
The evolutionary models indicate that J1741 is a $1.6 \pm 0.1$ Gyr old,
0.168(5)\msun, and $M_g=8.3$ mag WD, at a distance of $970 \pm 45$ pc. The
systematic uncertainties in these mass estimates are likely around
10\%, or 0.02\msun.

\subsection{Optical Photometry}

We obtained high-speed photometry of J0751 from two sites between 2012 February and 2013 April.
We obtained 45.3 h (5-30 s exposures) and 17.9 h (30-45 s exposures) of data
using the Argos \citep{nather04} and the Puoko-nui \citep{chote13} instruments on the McDonald
Observatory 2.1m and the Mt. John Observatory 1.0 m telescopes, respectively. 
All observations were obtained through a BG40 filter.
\citet{hermes12} obtained 9.5 hr of photometry for J1741 using Argos on the McDonald 2.1m
telescope in 2011 May and September. We obtained an additional 3.5 hr of high-speed photometry with
15-30 s exposures and the same setup in 2012 June and July. 
Our data reduction procedures are described in \citet{hermes12}.

\subsection{X-ray Observations}

If the companions to J0751 and J1741 are neutron stars (NS), they would have been spun up to
millisecond periods during the primary's red giant phase. Such millisecond pulsars (MSPs) 
would be detected in X-rays even if they were radio-quiet or if their pulsar beams were
missing our line of sight, due to gravitational bending of the X-rays originating from the
surface \citep{beloborodov02}. We are motivated by the X-ray detection of all known MSPs in
the globular cluster 47 Tuc \citep{heinke05}, allowing predictions of the X-ray emission
of other MSPs.
We observed both J0751 and J1741 with Chandra's ACIS-S detector in Very Faint mode, 
for 4.0 (on 2012 Dec 22) and 6.0 (on 2013 Jan 18) ksec, respectively.
We reprocessed the raw Chandra data using {\sevensize CIAO 4.5}.
Inspection of the 0.3-7 keV images showed no counts within 1$\arcsec$ of the position of either WD
(well beyond the 0.6$\arcsec$ 90\% confidence region for Chandra astrometry), and only 1 or 0 counts
within 5$\arcsec$ of J0751 or J1741, respectively.

We also obtained a 4.8 ksec XMM observation of J1741 on 2012 Oct 25.
We analyzed the standard pipeline products using {\sevensize SAS v13.0.0}.  Although the background was variable, we did not
reject any times based on high background, leaving us with 4.8, 4.7, and 3.2 ks of good time for the MOS1, MOS2,
and pn detectors, respectively. We filtered the data using standard filters for soft data,
and selecting the 0.2-1.5 keV energy range for the most effective
selection of soft sources. No source is visible within $1\arcmin$ of J1741 in any camera. We measure the counts within
an $8\arcsec$ region around J1741's position, which would contain 50\% of the 1.5 keV photons from a source at that position.
Including background subtraction, we find 2$\sigma$ upper limits of 15, 5 and 5 counts from J1741 in the pn, MOS1 and MOS2 error circles.

To infer upper limits on the X-ray flux of our targets, we first use the {\sevensize COLDEN} tool to interpolate the
\citet{dickey90} \ion{H}{i} survey, and estimate $N_H$ values of $5.9\times10^{20}$ cm$^{-2}$ and $3.7\times10^{20}$
cm$^{-2}$ towards J0751 and J1741 respectively. Since both targets are well above the Galaxy's \ion{H}{i} disk
(0.15 kpc scale height, e.g. \citealt{kalberla09}), we assume the full Galactic extinction.
We use the {\sevensize PIMMS} tool to compute the
unabsorbed 0.3-8 keV flux for a 134 eV blackbody (appropriate for the faintest 47 Tuc MSP).
We calculate 95\% confidence upper limits to the 0.3-8 keV X-ray fluxes, and thus to the 0.3-8 keV luminosities.
We calculate upper limits on the true countrate, using the observed countrate and Poisson statistics
\citep{gehrels86}; so for the Chandra observations, we used upper limits of 3 counts, corresponding to 95\%
confidence when 0 counts are observed. This gives $F_X$ upper limits of $3.7\times10^{-15}$ erg cm$^{-2}$ s$^{-1}$
for J1741 and $6.5\times10^{-15}$ erg cm$^{-2}$ s$^{-1}$ for J0751. We then use the 2$\sigma$ upper limits on
the distances (i.e. 1060 pc for J1741 and 1968 pc for J0751, see \S 2.1) to find upper limits of
$L_X<3.0\times10^{30}$ and $5\times10^{29}$ erg s$^{-1}$ for J0751 and J1741,
respectively. We confirm our result for J1741 using the XMM observation, which gives an upper limit of
$L_X<1.7\times10^{30}$ erg s$^{-1}$ from each of the pn and (combined) MOS cameras.

For J1741, our best 95\% confidence upper $L_X$ limit of $5\times10^{29}$ erg s$^{-1}$ is well below the luminosity of
any MSP in 47 Tuc, using the blackbody fluxes reported in \citet{bogdanov06} and a 4.5 kpc distance \citep[][2010 revision]{harris96}.
For J0751, however, our new estimate of a higher distance means that our requested observing time was insufficient to completely exclude MSP fluxes, giving an $L_X$ upper limit above six (of nineteen) of the 47 Tuc MSPs.
We can also select a $1\sigma$ upper limit (using an upper limit of 1.8 counts, and increasing the best-fit distance
by only 1$\sigma$), finding $L_X$(0.3-8 keV)$<1.67\times10^{30}$ erg s$^{-1}$. This is fainter than all but one (of 19)
MSPs in 47 Tuc.  
We conclude that our X-ray observations decisively reject a NS companion for J1741, and present strong
(but not conclusive) evidence against a NS companion for J0751. A Green Bank Telescope radio search has also failed to detect
MSP companions in both systems (Andrews 2013, priv. comm.).

\vspace{-0.2in}
\section{Binary Parameters}

\subsection{J0751}

\begin{figure}
\begin{center}
\includegraphics[width=2.8in,angle=0]{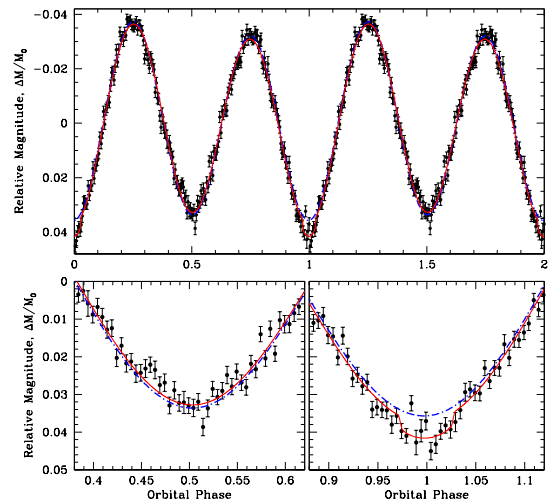}
\caption{High speed photometry of J0751 folded at the orbital period, binned into 200 phase bins, and
repeated for clarity. The dotted blue line shows a model fit including ellipsoidal variations and Doppler
beaming. The solid red line shows the best-fit model including
primary eclipses.
\label{fig:photj0751}}
\end{center}
\vspace{-0.1in}
\end{figure}

J0751 shows $K = 432.6 \pm 2.3$ \kms\ velocity variations
with an orbital period of $0.08001 \pm 0.00279$ d \citep{brown13}. 
Figure \ref{fig:photj0751} shows J0751's optical light curve.
A Fourier transform of our photometric dataset shows a well-resolved peak at 57.60907(3) min
due to $3.20 \pm 0.11$\% ellipsoidal variations. We use this signal to improve our estimate of the orbital period
to 0.08001260(4) d. 
The revised mass function
is $f= 0.671 \pm 0.011$ \msun, which yields a minimum companion mass of $0.97 \pm 0.01$\msun.
The non-detection of a NS companion constrains the inclination to $i\geq58.6^\circ$.
We detect a slight asymmetry
in the maxima of the folded light curve due to a $0.24 \pm 0.11$\% Doppler beaming signal \citep{zucker07}.
We also see substantially depressed minima at Phase 1 that do not appear fully represented by a
model of ellipsoidal variations and Doppler beaming. These features are evidence of eclipses of the low-mass
primary, which makes J0751 the fifth eclipsing double WD known and provides further
constraints on the system parameters. 
We use {\sevensize JKTEBOP} \citep{southworth05} and the limb-darkening coefficients of \citet{gianninas13}
to model this system, although our results are highly uncertain due to the relatively large
scatter of the data at Phase 1.
The best-fit model has $i=85.4^{+4.2}_{-9.4}$ deg, $R_1=0.155 \pm 0.020$\rsun, $R_2=0.0092 \pm 0.0026$\rsun,
and $M_2=0.97^{+0.06}_{-0.01}$\msun. The errors in these parameters are calculated from $10^5$ Monte-Carlo
simulations and represent 95.5\% (2$\sigma$) errors. Based on the $\log{g}$ measurement and inferred mass,
the radius should be $R_1 = 0.124 \pm 0.014$\rsun, which agrees reasonably with our light curve analysis.
Additional observations of the eclipses will be useful for more precise constraints on this system. 

\subsection{J1741}

\begin{figure}
\begin{center}
\includegraphics[width=2.8in,angle=0]{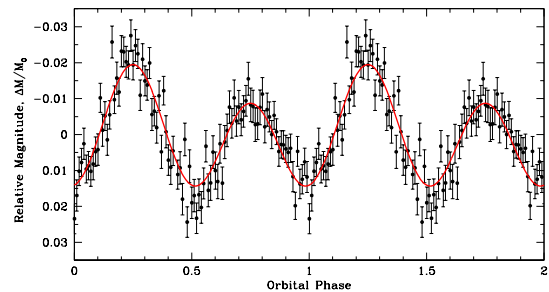}
\caption{High speed photometry of J1741, folded at the orbital period and
repeated for clarity. The red line shows the best-fit model including ellipsoidal
variations and Doppler beaming.
\label{fig:photj1741}}
\end{center}
\vspace{-0.1in}
\end{figure}

J1741 shows $K = 508 \pm 4$\kms\ velocity variations with an orbital period of
$P = 0.06111 \pm 0.00001$ \citep{brown12}. The mass function is
$f= 0.830 \pm 0.018$ \msun, and the minimum companion mass is $1.10 \pm 0.01$\msun.
The lack of a NS companion requires $i\geq64.9^\circ$.
Figure \ref{fig:photj1741} shows the optical light curve of J1741,
which exhibits signiﬁcant evidence for both Doppler beaming ($0.50 \pm 0.08$\%)
and ellipsoidal variations ($1.30 \pm  0.08$\%). \citet{hermes12} noted that
they might have also detected a reflection effect at the $2\sigma$ level, but our additional data
demonstrate that there is no significant reflection effect in this system.

There are no signs of eclipses in our light curve at the $4\sigma$ level
of 0.32\%. This constrains the inclination to $i\leq84.4^\circ$ and $M_2\geq1.11$\msun.
Based on the uncertainties in the spectroscopic surface gravity and derived mass,
we expect the primary to have a radius of $R_1 = 0.067 \pm 0.005$\rsun.
The amplitude of the ellipsoidal variations constrain $R_1$ to 0.069-0.079\rsun\
over the allowed inclinations of $i=64.9^{\circ}-84.4^{\circ}$.
Both values are in good agreement.

\vspace{-0.2in}
\section{The Future}

\subsection{AM CVn}

Our optical and X-ray observations demonstrate that
J0751 is a $0.19 \pm 0.02$\msun\ WD with a $0.97^{+0.06}_{-0.01}$\msun\ WD companion in a $P=115$ min
binary. Similarly, J1741 is a $0.17 \pm 0.02$\msun\  WD with an unseen $M\geq1.11$\msun\ WD companion in a $P=88$ min binary.
Assuming energy and angular momentum loss due to gravitational wave
radiation, the ELM WDs in J0751 and J1741 will reach their Roche lobe radii and start mass
transfer in about 270 and 160 Myr, respectively.

\begin{figure}
\hspace{0.1in}
\includegraphics[width=2.3in,angle=-90]{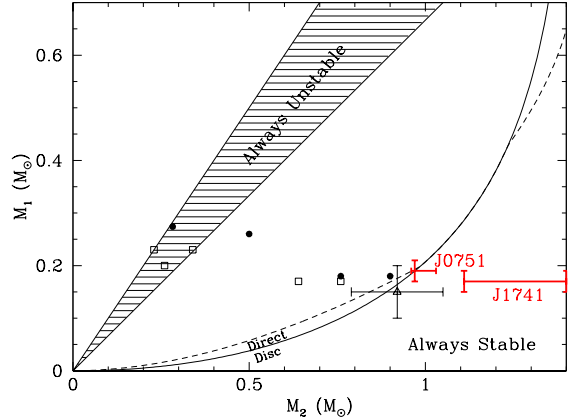}
\caption{Mass transfer stability for double WDs \citep{marsh04}. Disk accretion occurs in
the region below the solid line. J1741 (J0751) is clearly (likely) in this parameter range;
it will evolve into a stable mass transfer AM CVn.
Eclipsing double WD systems (filled circles), other ELM WD merger systems with
X-day data (open squares), and double-lined binary WD system J1257+5428 (open triangle) are also shown.
\label{fig:stab}}
\vspace{-0.1in}
\end{figure}

The stability of the mass transfer between double WDs depends on the mass ratio and the degree
of the spin-orbit coupling. Figure \ref{fig:stab} shows the dynamical stability
limit of \citet{marsh04} for different primary and secondary mass WDs. For
$q = M_2/M_1 > 2/3$, the mass transfer is unstable, leading to merger. For small q ($\leq0.2$),
stable mass transfer occurs through direct impact or disc accretion. For the
intermediate cases, the stability depends on the synchronization timescale
of the system. J1741 is clearly in the parameter range for stable mass transfer
through disc accretion. It will reach contact in $\sim10^8$ yr and is destined to
become a stable mass-transfer AM CVn system. This is the first identified progenitor of AM CVn.
With a mass ratio of $q=0.2$, J0751 may also turn into an AM CVn. However, the stability
of mass transfer and the future of J0751 are uncertain due to the unknown synchronization timescale.

Figure \ref{fig:stab} includes nine other double WD systems with mass constraints from
X-ray observations or eclipses. There are five ELM WD merger systems with X-ray data \citep{kilic11,kilic12},
and none show evidence of a NS companion. The lower-limit on the companion mass for each of
these five objects are shown as open squares. Given the unknown inclination and thus unknown companion mass, the future
of these objects is unclear. Out of the four eclipsing double WD systems previously known
\citep{steinfadt10,brown11a,vennes11,parsons11}, two have $q>0.5$; these will likely merge.
The future of the remaining two systems depends on the degree of the spin-orbit coupling.
Figure \ref{fig:stab} includes another potential AM CVn progenitor,
the double-lined spectroscopic binary SDSS J1257+5428 \citep{kulkarni10,marsh11}. Given the
difficulties in disentangling the spectra of the individual WDs in this system, the mass estimates
are uncertain; $M_1=0.15 \pm 0.05$\msun, $M_2=0.92 \pm 0.13$\msun. Hence,
the future of this binary is also unsettled. Figure \ref{fig:stab} shows that J1741 ($q=$ 0.12-0.15)
is unique as the first confirmed progenitor of stable-mass transfer AM CVn.

\citet{roelofs07} used six AM CVn found in the SDSS spectroscopic database to estimate
an AM CVn space density of $1.5 \times 10^{-6}$ pc$^{-3}$. After an extended search
for AM CVn in the SDSS, \citet{carter13} revised the space density estimate to
$5 \pm 3 \times 10^{-7}$ pc$^{-3}$, a factor of three smaller \citep[see][]{nissanke12}.
Using the former estimate,
\citet{brown11b} demonstrated that ELM WD binaries can explain 2-4\% of the AM CVn birthrate.
Given the lower space density of AM CVn and the larger number of ELM WDs binaries
now known, ELM WDs with massive companions can explain a significant fraction of AM CVn.

\subsection{SNe .Ia or Ia}

Recent studies by \citet{shen09} and \citet{shen10} confirm that AM CVn with $\geq0.8$\msun\ accretors will
achieve dynamical burning with envelope masses of $<0.1$\msun. Here, we have uncovered at least one (and perhaps
two) binary WD that should explode as SN .Ia in the next several hundred Myr.
Both J0751 and J1741 contain ELM WDs with $M>0.9$\msun\ companions. 
J1741 will start stable mass transfer in $\approx$160 Myr and will likely have a faint, thermonuclear
explosion within $10^8$ yr after that. Depending on the degree of spin-orbit coupling and the stability
of mass transfer, J0751 may follow a similar evolutionary path.

The envelope masses of $0.1$\msun\ are too small
to be relevant for the double-detonation models of \citet{nomoto82}, \citet{livne90}, and \citet{woosley94}.
However, \citet{fink10} find it inevitable that core detonation occurs after a helium shell detonation
on 0.94-1.39\msun\ CO WDs with shell masses of 0.0035-0.126\msun. \citet{shen13} also support the
double-detonation scenario for AM CVn; they find that helium shell detonations create converging shocks in the
core. These shocks are strong enough to ignite carbon and create SNe Ia events \citep[also see][]{moll13}.
This scenario essentially turns the SNe .Ia candidates into Type Ia candidates, and could explain
a fraction of thermonuclear SNe \citep{ruiter11}.
\citet{pakmor13} argue that the helium-ignited mergers of CO+CO and CO+He binary WDs
can explain the observed diversity of SNe Ia. In their model, a thin helium shell is ignited
even in the CO+CO WD merger, explaining normal and brighter Ia, whereas CO+He binary WDs
may lead to fainter SNe due to lower ejecta mass.

A remaining problem with the double-detonation model for AM CVn is that the explosion models
can match the light curves of SNe Ia but not their colors and spectra \citep{kromer10}. This is mostly due
to the effects of helium burning on the resulting spectra. However,
a good agreement between sub-Chandrasekhear-mass explosions and SNe Ia may still be feasible
depending on the initial composition of the helium shell. \citet{townsley12} demonstrate that
if post-shock radial expansion of the helium layer is taken into account, the predicted colors and
spectral features would be consistent with normal SNe Ia, except when the effects of the
extremely small helium shell is seen at earliest times.

An alternative endpoint for AM CVn involves core detonation, if the CO WD is already close to the
Chandrasekhar mass limit.
This channel likely contributes $\leq$1\% to the SNe Ia rate \citep{solheim05}.
The combined mass of the primary and secondary WDs in J0751 is $\approx$1.16\msun. On the other hand,
J1741's companion may be massive enough to reach the Chandrasekhar mass limit through accretion
from the ELM WD. Unfortunately, the core composition of the companions in these
two systems is not known. If they have ONe cores, then the double-detonation
scenario is highly unlikely as it is extremely difficult to ignite O-burning \citep{shen13}. In addition, accretion
from the companion may instead lead to an accretion-induced-collapse to a NS \citep{nomoto91}.

\vspace{-0.2in}
\section{Conclusions}

Direct discovery of the detached progenitors of AM CVn stars as well as SNe .Ia and Ia has so far been elusive.
Recent theoretical studies suggest
that sub-Chandrasekhar mass WDs in accreting systems, including AM CVn, may create thermonuclear SNe.
Ignition of a thin helium layer on an accreting CO WD may lead to .Ia explosions, and potentially SNe Ia.
Regardless of the theoretical problems in our understanding of the final outcome of helium-shell
detonations, it is clear that AM CVn with $M\geq0.8$\msun\ accretors are extremely interesting.
However, the component masses in AM CVn are rarely known and highly uncertain.

In this paper, we present the first confirmed progenitor(s) of AM CVn stars.
J0751 and J1741 are short period binary systems
containing 0.17-0.19\msun\ ELM WDs and $M>0.9$\msun\ WD companions. 
Due to its extreme mass ratio, J1741 will start stable mass
transfer and likely explode as SN .Ia in the next several hundred Myr. 
With $q=0.2$, J0751 is also a likely AM CVn and SN .Ia progenitor.
Depending on the companion masses and the core composition,
these systems may also detonate as SNe Ia. One has to wait only $\sim10^8$ yr to know the answer.

\vspace{-0.2in}
\section*{Acknowledgements}
MK and JJH acknowledge the support of the NSF under grants AST-1312678
and AST-0909107, and the Norman Hackerman Advanced Research Program.
COH acknowledges support from an NSERC Discovery Grant and an Ingenuity New Faculty Award.
DJS and PC thank the NZ Marsden Fund for financial support.

\noindent {\it Facilities: CXO, XMM, McDonald 2.1m, Mt. John 1.0m}

\end{document}